\documentstyle{article}
\begin{document}
\begin{center}
{\Large\bf Gravitational fields with a non Abelian\\
bidimensional Lie algebra of
symmetries}
\bigskip
\par 
{\bf G. Sparano $^{\#,\S }$, \ G. Vilasi $^{\#,\S \S }$ and A. M. 
Vinogradov 
$^{\#,\S }$}
\bigskip
\par 
{\it $^{\#}$ Istituto Nazionale di Fisica Nucleare, Gruppo
Collegato di Salerno, Italy.\\
$^{\S }$ Dipartimento di Matematica e Informatica, Universit\`{a} di
Salerno, Italy. \\
$^{\S \S }$ Dipartimento di Fisica ''E.R.Caianiello'', Universit\`{a} di
Salerno, Italy.}
\par
\begin{abstract}
Vacuum gravitational fields invariant for a bidimensional non Abelian 
Lie
algebra of Killing fields, are explicitly described. They are 
parameterized
either by solutions of a transcendental equation (the {\it tortoise 
equation}%
) or by solutions of a linear second order differential equation on the
plane. Gravitational fields determined {\it via} the tortoise equation, 
are
invariant for a $3$-dimensional Lie algebra of Killing fields with
bidimensional leaves. Global gravitational fields out of local ones are 
also
constructed.

\end{abstract}
\end{center}
In the last years a great deal of attention has been devoted to the
detection of gravitational waves. However, all the experimental devices,
interferometers or resonant antennas, are constructed coherently with
results obtained from the non covariant linearized Einstein field 
equations,
in close analogy with that is normally done in Maxwell theory of
electromagnetic fields.

Starting from the seventy's, however, new powerful mathematical methods 
have
been invented to deal with nonlinear evolution equations and their exact
solutions. One of this methods, namely a suitable generalization of the 
{\it %
\ \ Inverse Scattering Transform}, allowed to solve reduced vacuum 
Einstein
field equations and to obtain {\it solitary waves solutions} \cite{BZ78}
(see \cite{Ve93} and references therein).

This paper is the first in a series devoted to the study of 
gravitational
fields $g$ admitting a Lie algebra ${\cal G}$ of Killing fields. The 
case of
a non Abelian bidimensional Killing Lie algebra has been only partially
studied. Here, this case will be completely analyzed within the 
following
general problem which, as we will see, emerges naturally.

I. the distribution ${\cal D}$, generated by the vector fields of ${\cal 
G}$%
, is bidimensional.

II. the distribution ${\cal D}^{{\cal \perp }}$ orthogonal to ${\cal 
D}$, is
integrable and transversal to $D$.

According to whether $\dim {\cal G}$ is $2$ or $3$, two qualitatively
different cases occur.

A bidimensional ${\cal G}$, is either Abelian (${\cal A}_{2}$) or
non-Abelian (${\cal G}_{2}$). A metric $g$ satisfying I and II, with 
${\cal %
\ G }={\cal A}_{2}$ or ${\cal G}_{2}$, will be called ${\cal G}${\it \
-integrable} .

The study of ${\cal A}_{2}$-integrable Einstein metrics goes back to
Einstein and Rosen \cite{ER37}, Rosen \cite{Ro54}, Kompaneyets 
\cite{Ko58},
Geroch \cite{Ge72}, Belinsky and Khalatnikov \cite{BK70}.

The greater rigidity of ${\cal G}_{2}$-integrable metrics, for which 
some
partial results can be found in \cite{Ha88,AL92,Ch98}, allows an 
exhaustive
analysis. It will be shown that they are parameterized by solutions of a
linear second order differential equation on the plane which, in its 
turn,
depends linearly on the choice of a ${\bf j}${\it -harmonic} function 
(see
later). Thus, this class of solutions has a {\it bilinear structure} 
and, as
such, admits two{\it \ superposition laws. }

When $\dim $ ${\cal G}=3$, assumption II follows from I and the local
structure of this class of Einstein metrics can be explicitly described.
Some well known exact solutions \cite{Pe69,KSHM80}, {\it e.g.}
Schwarzschild, belong to this class.

Besides the new local ${\cal G}_{2}$-integrable solutions, a procedure 
to
construct new global solutions, suitable for all such ${\cal 
G}$-integrable
metrics, will be also described.

The following notation will be adopted

\noindent {\it Metric}: a non-degenerate symmetric $\left( 0,2\right) $
tensor field,

\noindent ${\cal K}il\left( g\right)$: the Lie algebra of all Killing 
fields
of a metric $g$,

\noindent {\it Killing algebra}: a sub-algebra of ${\cal K}il\left( 
g\right) 
$.

{\bf Semiadapted coordinates.} Let $g$ be a metric on the space-time $M$ 
(a
connected smooth manifold) and ${\cal G}_{2}$ one of its Killing 
algebras
whose generators $X,Y$ satisfy 
\begin{equation}
\lbrack X,Y]=sY,\ \ \ s\in {\cal R}
\end{equation}
The Frobenius distribution ${\cal D}$ generated by ${\cal G}_{2}$ is
bidimensional and a chart $(x^{1},\,x^{2},\,x^{3},\,x^{4})$ exists such 
that 
\begin{equation}
X=\frac{\partial }{\partial x^{3}},\,\,\ \,Y=\left( \exp sx^{3}\right) 
\frac{%
\partial }{\partial x^{4}}.
\end{equation}
>From now on such a chart will be called {\it semiadapted }(to the 
Killing
fields).

\paragraph{\it Invariant metrics}

It can be easily verified \cite{SVV00} that in a semiadapted chart $g$ 
has
the form 
\begin{eqnarray*}
g &=&g_{ij}dx^{i}dx^{j}+2\left( l_{i}+sm_{i}x^{4}\right)
\,dx^{i}dx^{3}-2m_{i}dx^{i}dx^{4}+ \\
&&\left( s^{2}\lambda \left( x^{4}\right) ^{2}-2s\mu x^{4}+\nu \right)
dx^{3}dx^{3}+ \\
&&2\left( \mu -s\lambda x^{4}\right) dx^{3}dx^{4}+\lambda 
dx^{4}dx^{4},\quad
i=1,2;j=1,2
\end{eqnarray*}
with\ $g_{ij}$, $m_{i}$, $l_{i}$, $\lambda $, $\mu $, $\nu $ arbitrary
functions of $\left( x^{1},x^{2}\right) $.

{\bf Killing leaves.} Condition II allows to construct semi-adapted 
charts,
with new coordinates $\left( x,y,x^{3},x^{4}\right) $, such that the 
fields $%
e_{1}=\frac{\partial }{\partial x}$, $e_{2}=\frac{\partial }{\partial 
y}$,
belong to ${\cal D}^{\bot }$. In such a chart, called from now on {\it %
adapted}, the components $l_{i}$'s and $m_{i}$'s vanish.

We will call {\it Killing leaf} an integral (bidimensional) submanifold 
of $%
{\cal D}$ and {\it orthogonal leaf } an integral (bidimensional) 
submanifold
of ${\cal D}^{\bot }$. Since ${\cal D}^{\bot }$ is transversal to ${\cal 
D}$%
, the restriction of $g$ to any Killing leaf, $S$, is non-degenerate. 
Thus, $%
\left( S,\left. g\right| _{S}\right) $ is a homogeneous bidimensional
Riemannian manifold. Then, the Gauss curvature $K\left( S\right) $ of 
the
Killing leaves is constant (depending on the leave). In the chart 
($p=\left.
x^{3}\right| _{S}$, $q=\left. x^{4}\right| _{S}$) one has 
\[
\left. g\right| _{S}=\left( s^{2}\widetilde{\lambda 
}q^{2}-2s\widetilde{\mu }%
q+\widetilde{\nu }\right) dp^{2}+2\left( \widetilde{\mu }-s\widetilde{%
\lambda }q\right) dpdq+\widetilde{\lambda }dq^{2}, 
\]
where $\widetilde{\lambda },\widetilde{\mu },\widetilde{\nu }$, being 
the
restrictions to $S$ of ${\lambda },{\mu },\nu $, are constants, and 
\begin{equation}
K\left( S\right) =\widetilde{\lambda }s^{2}\left( \widetilde{\mu }^{2}-%
\widetilde{\lambda }\widetilde{\nu }\right) ^{-1}.  \nonumber
\end{equation}

{\bf Einstein metrics when $g(Y,Y)\neq 0$.} In the considered class of
metrics, vacuum Einstein equations, $R_{\mu \nu }=0$, can be completely
solved \cite{SVV00}. If the Killing field $Y$ is not of {\it light 
type}, 
{\it i.e.} $g(Y,Y)\neq 0$, then in the adapted coordinates $\left(
x,y,p,q\right) $ the general solution is 
\begin{equation}
g=f(dx^{2}\pm dy^{2})+\beta
^{2}[(s^{2}k^{2}q^{2}-2slq+m)dp^{2}+2(l-skq)dpdq+kdq^{2}]  \label{snew}
\end{equation}
where $f=-\frac{1}{2s^{2}k}\bigtriangleup _{\pm }\beta ^{2}$, and$\ \ 
\beta
\left( x,y\right) $ is a solution of the {\it tortoise equation} 
\[
\beta +A\ln \left| \beta -A\right| =u\left( x,y\right) ,
\]
the function $u$ being a solution either of Laplace or d' Alembert 
equation, 
$\bigtriangleup _{\pm }u=0$, $\bigtriangleup _{\pm }=\partial 
_{xx}^{2}\pm
\partial _{yy}^{2}$, such that $\left( \partial _{x}u\right) ^{2}\pm 
\left(
\partial _{y}u\right) ^{2}\neq 0$. The constants $k,l,m$ are constrained 
by $%
km-l^{2}=\pm 1$, $k\neq 0$.

\paragraph{Canonical form of metrics when $g(Y,Y)\neq 0$}

The gauge freedom of the above solution, allowed by the function $u,$ 
can be
locally eliminated by introducing the coordinates $(u,v,p,q)$, the 
function $%
v(x,y)$ being conjugate to $u(x,y)$, {\it i.e.} $\bigtriangleup _{\pm 
}v=0$
and $u_{x}=v_{y},u_{y}=\mp v_{x}$. In these coordinates the metric $g$ 
takes
the form (local ''Birkhoff's theorem'')

\[
g=\frac{exp[{\frac{u-\beta }{A}}]}{2s^{2}k\beta }(du^{2}\pm 
dv^{2})+\beta
^{2}[(s^{2}k^{2}q^{2}-2slq+m)dp^{2}+2(l-skq)dpdq+kdq^{2}]
\]
with $\beta \left( u\right) $ a solution of $\beta +A\ln \left| \beta
-A\right| =u$.

\paragraph{Normal form of metrics when $g(Y,Y)\neq 0$.}

In {\it geographic coordinates} $\left( \vartheta ,\varphi \right) $ 
along
Killing leaves one has $\left. g\right| _{S}=\beta ^{2}\left[ d\vartheta
^{2}+{\cal F}\left( \vartheta \right) d\varphi ^{2}\right] $, where 
${\cal F}%
\left( \vartheta \right) $ is equal either to $\sinh ^{2}\vartheta $ or 
$%
-\cosh ^{2}\vartheta $, depending on the signature of the metric. Thus, 
in
the {\it normal coordinates}, $\left( r=2s^{2}k\beta ,\tau =v,\vartheta
,\varphi \right) $, the metric takes the form 
\begin{equation}
g=\varepsilon _{1}\left( \left[ 1-\frac{A}{r}\right] d\tau ^{2}\pm 
\left[ 1-%
\frac{A}{r}\right] ^{-1}dr^{2}\right) +\varepsilon _{2}r^{2}\left[
d\vartheta ^{2}+{\cal F}\left( \vartheta \right) d\varphi ^{2}\right] 
\label{nfs}
\end{equation}
where $\varepsilon _{1}=\pm 1$, $\varepsilon _{2}=\pm 1$ with a choice
coherent with the required {\it signature} $2$.

The geometric reason for this form is that, when $g(Y,Y)\neq 0$, a third
Killing field exists which together with $X$ and $Y$ constitute a basis 
of $%
so(2,1)$. The larger symmetry implies that the geodesic equations 
describe a 
{\it non-commutatively integrable system } \cite{SV00}, and the
corresponding geodesic flow projects on the geodesic flow of the metric
restricted to the Killing leaves. {\it The above local form does not 
allow,
however, to treat properly the singularities appearing inevitably in 
global
solutions.}{\em \ }The metrics (\ref{snew}), although they all are 
locally
diffeomorphic to (\ref{nfs}), play a relevant role in the construction 
of
new global solutions as described later.

{\bf Einstein metrics when $g(Y,Y)=0$.} If the Killing field $Y$ is of 
{\it %
\ \ light type}, then the general solution of vacuum Einstein equations, 
in
the adapted coordinates $\left( x,y,p,q\right) $, is given by 
\begin{equation}
g=2f(dx^{2}+dy^{2})+\mu [(w\left( x,y\right) -2sq)dp^{2}+2dpdq],  
\label{new}
\end{equation}
where $\mu =D\Phi +B$; $D,B\in {\cal R}$, $\Phi $ is a non constant 
harmonic
function, $f=\pm \left( \nabla \Phi \right) ^{2}/\sqrt{\left| \mu 
\right| }$
, and $w\left( x,y\right) $ is a solution of 
\[
\Delta w+\left( \partial _{x}\ln \left| \mu \right| \right) \partial
_{x}w+\left( \partial _{y}\ln \left| \mu \right| \right) \partial 
_{y}w=0. 
\]

Special solutions are $w=\widetilde{\mu }$, $\,\,w=\ln \left| \mu 
\right| ,$
where $\widetilde{\mu }$ is the harmonic function conjugate to $\mu $. 
When $%
\mu $ is not constant, in the coordinates $\xi =\mu +\widetilde{\mu },$ 
$%
\eta =\mu -\widetilde{\mu }$, the above equation appears to be the {\it 
\
Darboux equation} 
\[
\left( \xi +\eta \right) \left( \partial _{\xi \xi }^{2}+\partial _{\eta
\eta }^{2}\right) w+\partial _{\xi }w+\partial _{\eta }w=0. 
\]

In this case the curvature of Killing leaves vanishes.

The new solutions (\ref{new}) together with (\ref{snew}) exhaust all 
local
Lorentzian Ricci-flat metrics invariant for a ${\cal G}_2$ Lie algebra.

{\bf Ricci-flat $g$ with $\dim {\cal K}$}${\it il}${\bf $\left( \left.
g\right| _{S}\right) =3$ \& $\dim {\cal S}=2$.} In view of the 
construction
of global solutions, the previous results suggest to consider with the 
same
approach all metrics having 3-dimensional Killing algebras with
bidimensional leaves. A Killing algebra ${\cal G}$ of a metric $g$ will 
be
called{\it \ normal }if the restrictions of $g$ to Killing leaves $S$ of 
$%
{\cal G}$ are non-degenerate. Obviously, a normal Killing algebra ${\cal 
G}$
is isomorphic to a subalgebra of ${\cal K}il\left( \left. g\right|
_{S}\right) $. Thus, when $\dim {\cal G}=3$ and the Killing leaves are
bidimensional, ${\cal G}={\cal K}il\left( \left. g\right| _{S}\right) $. 
In
this situation there are just five options for ${\cal K}il\left( \left.
g\right| _{S}\right) $ and therefore for ${\cal G}$ : 
\[
so\left( 2,1\right) ,\,{\cal K}il\left( dp^{2}-dq^{2}\right) 
,\,\,so\left(
3\right) ,\,{\cal K}il\left( dp^{2}+dq^{2}\right) ,\,\ {\cal A}_{3}\,. 
\]

The method used before allows to describe completely Einstein metrics
admitting one of these algebra.

The gravitational fields invariant for $so(2,1)$ and ${\cal K}%
il(dp^{2}-dq^{2})$, which are the only ones possessing non Abelian
bidimensional subalgebras, can be found among solutions (\ref{snew}) and 
(%
\ref{new}).

The gravitational fields invariant for $so(3)$ and corresponding to a
positive choice of the solution $\beta (u)$ of the tortoise equation 
have
the following local form: 
\begin{equation}
g=\bigtriangleup _{\pm }{\beta }^{2}(dx^{2}\pm dy^{2})+{\beta 
}^{2}\left[
d\vartheta ^{2}+\sin ^{2}\vartheta d\varphi ^{2}\right] .
\end{equation}
The choice of normal coordinates as in equation (\ref{nfs}) and of minus 
sign gives
the Schwarzschild solution with a new insight to the physical meaning of 
the
so called {\it Regge-Wheeler tortoise coordinate }\cite{Wa84}.

The gravitational fields $g$ invariant for ${\cal K}il(dp^{2}+dq^{2})$, 
have
the local form

\begin{eqnarray*}
g &=&2f\left( dx^{2}-dy^{2}\right) +\alpha \left( x,y\right) \left[
dr^{2}+r^{2}d\varphi ^{2}\right] , \\
\alpha &\equiv &C_{1}F\left( x+y\right) +C_{2}G\left( x-y\right)
+C,\,\,\,f\equiv F^{\prime }G^{\prime }/\sqrt{\left| \alpha \right| },
\end{eqnarray*}
$F$ and $G$ being arbitrary functions, $C$, $C_{1}$, $C_{2}$, arbitrary
constants such that $\alpha $ and $f$ are nonvanishing.

The Lie algebra ${\cal A}_{3}$ belongs to the Abelian case of 
\cite{BZ78}.

{\bf Global solutions.} Any of previous metrics is fixed by a solution 
of
the wave or Laplace equation, and a choice

\begin{itemize}
\item  of the constant $A$ and one of the branches of a solution of the
tortoise equation, if $g(X,Y)\neq 0$.

\item  of a solution of Darboux equation, if $g(X,Y)=0$.
\end{itemize}

The metric manifold $\left( M,g\right) $ has a bundle structure whose 
fibers
are the Killing leaves and whose base ${\cal W}$ is a bidimensional 
manifold
diffeomorphic to the orthogonal leaves. The wave and Laplace equations
mentioned above are defined on ${\cal W}$. Thus, the problem of the
extension of our local solutions is reduced to that of the extension of 
$%
{\cal W}$. \ Such an extension carries a geometric structure, the ${\bf 
j}$ 
{\it -complex structure}, that gives an intrinsic sense to the notion of 
the
wave and Laplace equations and clarifies what variety of different
geometries is, in fact, obtained.

\paragraph{${\bf j}$-complex structures}

In full parallel with ordinary complex numbers, ${\bf j}${\it -complex
numbers }of the form $z=x+{\bf j}y$, with ${\bf j}^{2}=-1,0,1$, can be
introduced. Thus, a ${\bf j}${\it -complex analysis }can be developed by
defining ${\bf j}${\it -holomorphic} functions as ${\cal R}_{{\bf 
j}}^{2}$
-valued differentiable functions of $z=x+{\bf j}y$, where ${\cal 
R}_{{\bf j}%
}^{2}$ is the algebra of ${\bf j}$-complex numbers. Just as in the case 
of
ordinary complex numbers, a function $f\left( z\right) =u\left( 
x,y\right) + 
{\bf j}v\left( x,y\right) $ is ${\bf j}$-{\it holomorphic\ }if and only 
if
the ${\bf j}${\it -Cauchy-Riemann} conditions hold: 
\begin{equation}
u_{x}=v_{y},\qquad u_{y}={\bf j}^{2}v_{x}
\end{equation}
{\it \ }The compatibility conditions of the above system require that 
both $%
u $ and $v$ satisfy the ${\bf j}${\it -Laplace equation}, that is 
\begin{equation}
-{\bf j}^{2}u_{xx}+u_{yy}=0,\qquad -{\bf j}^{2}v_{xx}+v_{yy}=0.
\end{equation}
Of course, the ${\bf j}${\it -}Laplace{\it \ }equation reduces for ${\bf 
j}
^{2}=-1$ to the ordinary Laplace equation, while for ${\bf j}^{2}=1$ to 
the
wave equation.

A bidimensional manifold ${\cal W}$ supplied with a ${\bf j}$-{\it 
complex
structure} is called a ${\bf j}$-{\it complex curve}. Obviously, for 
${\bf j}
^{2}=-1$ a ${\bf j}$-complex curve{\it \ } is just a $1$-dimensional 
complex
manifold. The case ${\bf j}^{2}=0$ will not be considered.

Thus, any global metric is associated with a pair consisting of a ${\bf 
j}$
-complex curve ${\cal W}$ and a ${\bf j}$ -harmonic function $u$ on it.

\paragraph{Model solutions}

The pairs $\left( {\cal W},u\right) $ and $\left( {\cal W}^{\prime
},u^{\prime }\right) $, corresponding to two equivalent solutions, are
related by an invertible ${\bf j}$-holomorphic map $\Phi :\left( {\cal 
W}
,u\right) \rightarrow \left( {\cal W}^{\prime },u^{\prime }\right)$ such
that $\Phi ^{\ast }\left( u^{\prime }\right) =u$.

Particularly important are then the {\it model solutions}, namely those
solution for which $\left( {\cal W},u\right) =\left( {\cal R}_{{\bf j}%
}^{2},x\right) $. The pair $\left( {\cal R}_{{\bf j}.}^{2},x\right) $ is 
{\it universal } in the sense that any solution characterized by a given
pair $\left( {\cal W},u\right) $ is the pull-back of a model solution by 
a $%
{\bf j}$-holomorphic map $\Phi :{\cal W}\longrightarrow {\cal R}_{{\bf 
j}%
}^{2}$ defined uniquely by the relations $\Phi ^{\ast }\left( x\right) 
=u$
and $\Phi ^{\ast }\left( y\right) =v$, where $v$ is conjugated with $u$.

It will be now described in detail how to construct global solutions in 
the
case $\dim {\cal K}il\left( g_{\Sigma }\right) =3$. The remaining cases 
can
be found in \cite{SVV00}.

Let us first consider $so\left( 3\right) $ and $so\left( 2,1\right) $.
Denote by $\left( \Sigma ,g_{\Sigma }\right) $ a homogeneous 
bidimensional
Riemannian manifold, whose Gauss curvature $K\left( g_{\Sigma }\right) 
$, if
different from zero, is normalized to $\pm 1$. Let $\left( {\cal 
W},u\right) 
$ be a pair consisting of a ${\bf j}$-complex curve ${\cal W}$ and a 
${\bf j}
$-harmonic function $u$ on ${\cal W}$. The bundle structure $\pi
_{1}:M\rightarrow {\cal W}$ canonically splits in the product ${\cal W}
\times \Sigma $. Denote by $\pi _{2}:M\rightarrow \Sigma $ the also 
natural
projection of $M={\cal W}\times \Sigma $ on $\Sigma $. Then, the above 
data
determine the following Ricci-flat manifold $\left( M,g\right) $ with 
\begin{equation}
g=\pi _{1}^{\ast }\left( g_{\left[ u\right] }\right) +\pi _{1}^{\ast 
}\left(
\beta ^{2}\right) \pi _{2}^{\ast }\left( g_{\Sigma }\right)
\end{equation}
where $\beta \left( u\right) $ is implicitly determined by the tortoise
equation, and 
\begin{equation}
g_{\left[ u\right] }=\pm \frac{(\beta -A)}{\beta }\left( du^{2}-{\bf j}%
^{2}dv^{2}\right) .
\end{equation}

In the case of normal Killing algebras isomorphic to ${\cal K}il\left(
dx^{2}\pm dy^{2}\right) $ it is sufficient to consider Ricci-flat 
manifolds $%
M$ of the form 
\begin{equation}
M={\cal W}\times \Sigma ,\qquad g=\pi _{1}^{\ast }\left( g_{\left[ 
u\right]
}\right) +\pi _{1}^{\ast }\left( u\right) \pi _{2}^{\ast }\left( 
g_{\Sigma
}\right)
\end{equation}
where $\left( \Sigma ,g_{\Sigma }\right) $ is a flat bidimensional 
manifold
and 
\begin{equation}
g_{\left[ u\right] }=\pm \frac{1}{\sqrt{u}}\left( du^{2}-{\bf j}%
^{2}dv^{2}\right) .
\end{equation}

{\bf Examples}

\paragraph{Algebraic solutions}

Let ${\cal W}$ be an algebraic curve over ${\cal C}$, understood as a 
${\bf %
j }$-complex curve with ${\bf j}^{2}=-1$. With a given meromorphic 
function $%
\Phi $ on ${\cal W}$ a pair $\left( {\cal W}_{\Phi },u\right) $ is
associated, where ${\cal W}_{\Phi }$ is ${\cal W}$ deprived of the poles 
of $%
\Phi $ and $u$ the real part of $\Phi $. A solution (metric) constructed
over such a pair will be called algebraic. Algebraic metrics are 
generally
singular \cite{Vi87}, {\it e.g.} they are degenerate along the fiber 
$\pi
^{-1}\left( a\right) $ if at $a\in {\cal W}$ $d_{a}u=0$.

\paragraph{A star ''outside'' the universe}

\ The Schwarzschild solution describes a star generating a space around
itself. It is an $so\left( 3\right) $-invariant solution of the vacuum
Einstein equations. Its $so\left( 2,1\right) $ analogue shows a star
generating the space only on ''{\it one side of itself''}. More exactly, 
the
fact that the space in the Schwarzschild universe is formed by a $1$%
-parameter family of concentric spheres allows one to give a sense to 
the
adverb {\it around}. In the $so\left( 2,1\right) $ case the space is 
formed
by a $1$-parameter family of {\it concentric} hyperboloids. The 
adjective 
{\it concentric} means that the curves orthogonal to hyperboloids are
geodesics and metrically converge to a singular point. This explains in 
what
sense this singular point generates the space only on ''{\it one side of
itself''}.

The next example shows how the introduction of ${\bf j}$-complex 
structures
allows to manipulate already known Einstein metrics to get new ones also
with singularities.

\paragraph{The ''square root'' of the Schwarzschild universe}

It is an Einstein metric induced by the ${\bf j}${\it -quadratic map: 
}$%
z\longrightarrow \frac{{\bf j}}{2}z^{2}$, with ${\bf j}^{2}=1$, from a 
model
of the Schwarzschild type. The pair $\left( {\cal W},u\right) $ is now 
$%
\left( {\cal R}_{{\bf j}.}^{2},xy\right) $and the local expression of 
the
metric is 
\[
g=\frac{e^{\frac{\beta +xy}{A}}}{2\beta }\left( x^{2}-y^{2}\right) 
\left(
dy^{2}-dx^{2}\right) +\beta ^{2}\left[ d\vartheta ^{2}+{\cal F}\left(
\vartheta \right) d\varphi ^{2}\right] , 
\]
where ${\cal F}$, depending on the Gauss curvature $K$ of the Killing
leaves, is one of the functions $\sin ^{2}\vartheta ,$ $\sinh 
^{2}\vartheta $%
, $-\cosh ^{2}\vartheta $. The metric is degenerate along the lines 
$x=\pm y$%
. The Einstein manifolds so obtained consist of four regions soldered 
along
the degeneracy lines and could be interpreted \cite{SVV00} as 
''parallel''
universes generated by ''parallel'' stars. By repeating this procedure 
one
discover {\it foam-like universes.}

A detailed discussion of these and other new solutions will appear in a
forthcoming paper.

\vspace{0.1in}

\vspace{0in}\vspace{0in}G.S. and G.V. wish to thank G. Bimonte, B. 
Dubrovin,
and G. Marmo for their interest and remarks.


\begin{thebibliography}{99}
\bibitem{AL92}  B.N. Aliev and A.N.Leznov, J . Math. Phys. {\bf 33} n. 
7,
2567 (1992).

\bibitem{BK70}  V.A.Belinsky and I.M.Khalatnikov{\it ,} Sov. Phys. Jetp. 
{\bf 30}, 6 (1970).

\bibitem{BZ78}  V.A.Belinsky and V.E.Zakharov, Sov. Phys. Jetp {\bf 48}, 
6
(1978);{\bf \ 50}, 1 (1979).

\bibitem{Ch98}  F. J. Chinea, Class. Quantum Grav. {\bf 15}, 367 (1998).

\bibitem{ER37}  A.Einstein and N.Rosen, J.Franklin Inst. {\bf 223}, 43
(1937).

\bibitem{Ge72}  R.Geroch,{\it \ }J.Math.Phys. {\bf 13}, 3 (1972).

\bibitem{Ha88}  M.Hallisoy,{\it \ }J.Math.Phys. {\bf 29}, 2 (1988).

\bibitem{Ko58}  A.S.Kompaneyets, Sov. Phys. JETP {\bf 7}, 659 (1958).

\bibitem{KSHM80}  D.Kramer, H.Stephani, E.Herlt, M.MacCallum, {\it Exact
solutions of Einstein field equations }(Cambridge University Press,
Cambridge 1980).

\bibitem{Pe69}  A.Z.Petrov, {\it Einstein spaces, }(Pergamon Press,{\it 
\ }
New York, 1969).

\bibitem{Ro54}  N.Rosen, Bull. Res. Coun. Isr. {\bf 3}, 328 (1954).

\bibitem{SV00}  G.Sparano and G.Vilasi, J. Geom. Phys. {\bf 36}, 270 
(2000)

\bibitem{SVV00}  G.Sparano, G.Vilasi and A.M.Vinogradov, {\it Einstein
metrics with bidimensional Killing leaves} I and II, Dip. Fisica {\it 
ERC}
DFC-1-01; DFC-2-01 to be published.

\bibitem{Ve93}  E.Verdaguer, Phys. Rep. {\bf 229}, n. 1\&2, 1 (1993).

\bibitem{Vi87}  A.M.Vinogradov, {\it Geometric singularities of 
solutions of
partial differential equations}, Proc. Conf. Diff. Geom. and Appl. 
(Reidel,
Dordrecht 1987).

\bibitem{Wa84}  R.Wald, {\it General Relativity} (University of Chicago
Press, Chicago, 1984)
\end{thebibliography}
\end{document}